\documentclass[a4paper,12pt]{article}
\usepackage{graphicx}
\usepackage{color}
\usepackage{cite}
\usepackage{epsfig}
\usepackage{amsfonts}
\title{Thermodynamic Geometry: \\
\vspace{.2cm} Evolution, Correlation and Phase Transition}
\author{}
\vspace{-3cm}
\date{ S. Bellucci$^{1,} $ \thanks{\noindent bellucci@lnf.infn.it}\ \ and
 B. N. Tiwari$^{1,}$ \thanks{\noindent tiwari@lnf.infn.it}\\
\vspace{0.5cm}
$^{1}$ INFN-Laboratori Nazionali di Frascati\\
Via E. Fermi, 40 -- I - 00044 Frascati\\
Rome, Italy.}
\begin{document}
\maketitle
\begin{abstract}
Under the fluctuation of the electric charge and atomic mass, this paper considers
the theory of the thin film depletion layer formation of an ensemble of finitely excited,
non-empty $d/f$-orbital heavy materials, from the thermodynamic geometric perspective.
At each state of the local adiabatic evolutions, we examine the nature of the thermodynamic parameters,
\textit{viz.}, electric charge and  mass, changing at each respective embeddings.
The definition of the intrinsic Riemannian geometry and differential topology offers
the properties of (i) local heat capacities, (ii) global stability criterion and (iv) global
correlation length. Under the Gaussian fluctuations, such an intrinsic geometric consideration
is anticipated to be useful in the statistical coating of the thin film layer of a desired quality-fine
high cost material on a low cost durable coatant.
From the perspective of the daily-life applications, the thermodynamic geometry is thus
intrinsically self-consistent with the theory of the local and global economic optimizations.
Following the above procedure, the quality of the thin layer depletion could  self-consistently
be examined to produce an economic, quality products at a desired economic value.
\end{abstract}
\vspace{0.9cm}
\newpage
\section{Introduction}
Thermodynamic geometry has a wide class of applications in the
domain of the statistical mechanics and black hole physics. From
the physical fronts of the intrinsic Riemannian geometry, the
motivational bootstrapping fundamentals were introduced by
Wienhold \cite{wien1,wien2} and Ruppenier \cite{RuppeinerA20}, as
early as the 1975. Wienhold has introduced the notion of the
thermodynamic geometry from the chemical perspective. Soon after
the initiation of Wienhold, Ruppenier revived the subject by
reformulating the Weinhold inner product structure in the entropy
representation, and thus the conformally related Ruppenier's
thermodynamic geometric description
\cite{RuppeinerA20,RuppeinerRMP,RuppeinerPRL,
RuppeinerA27,RuppeinerA41} for diverse condense matter
configurations. Specifically, Ruppenier has expanded the
applicability of the thermodynamic geometry, by extending it's
framework to the black hole solutions in Einstein's general
relativity \cite{rupgr} and thereby he showed that the black hole
solutions of the general relativity \cite{waldGR} are
thermodynamically unstable.

Ruppeiner has further shown that the notion of the thermodynamic fluctuation theory \cite{Huang,Landau},
in addition to the thermodynamic laws, allows a remarkable physical interpretation of the intrinsic
geometric structure in terms of the probability distribution of the fluctuations, and thus the relationship
of the thermodynamic scalar curvature with critical phenomena.
Aman, Bengtsson, Pidokrajt and Lozano-Tellechea \cite{0304015v1,0510139v3,Arcioni} have extended
the framework of Ruppenier's thermodynamic geometry for diverse four dimensional black holes. Thereby,
the nature of the associated thermodynamic configurations could be properly understood from the viewpoint
of the intrinsic thermodynamic geometry. Since a decade, there have been a large number of excitements
\cite{math-ph/0507026,cai1}, revealing the thermodynamic geometric properties of such black holes. Further
investigations \cite{callen,Tisza} revealed that the equilibrium thermodynamic systems possess interesting
geometric thermodynamic structures.

Recent studies of the thermodynamics of a class of black holes
have elucidated interesting aspects of the underlying phase
transitions and their relations with the moduli spaces of
$\mathcal N \geq 2$ supergravity compactifications and the quantum
mechanical investigations, in the context of extremal black holes
\cite{9707203v1,9702103,0209114,Bull1,Bull2,Bull3,0412322,FHM,SF1,SF2,SF3}.
Subsequently, for the extremal black holes in string theory, the
exact matchings between the macroscopic entropy and the
microscopic entropy have been obtained in the leading and
subleading orders in the large charge asymptotic expansion
\cite{attrac1,attrac2,attrac3,attrac4,attrac5,attrac6,attrac7}. In
order to establish a more general variational technique to compute
the higher derivative corrections to the thermodynamic quantities,
Sen \cite {Sen1,Sen2,Sen3,Sen4,Sen5,Sen6} led down an alternative
analysis involving a non-trivial adaptation of the Wald formalism
(offering a generally covariant higher derivative theories of
gravity \cite {wald1,wald2,wald3}). The attractor equations follow
from the extremization of the Sen entropy function, and thus the
understanding of the entropy as an attractor fixed point horizon
quantity for the charged (extremal) black holes. Typically, the
generalized entropy function formalism is mostly independent of
the supersymmetry considerations and thus a better applicability
for the (extremal) non-supersymmetric black holes
\cite{0611166,Sen7,Sen8,sen9,sen10,0611140}.

In this framework, Bellucci and Tiwari \cite{BNTBull} have extended the framework of the
thermodynamic geometry to the various higher dimensional black holes in the string theory
and M-theory. Their investigation shows that the higher dimensional black branes are generically
unstable from the viewpoint of the limiting Ruppenier's thermodynamics state-space manifolds.
The associated thermodynamic properties of the BTZ black holes \cite{SST} and leading order
extremal black holes \cite{0606084v1} explore the similar behavior. In the viewpoint of the stringy
$\alpha^{\prime}$ corrections, Tiwari \cite{bnt} has demonstrated that most of the extremal and
non-extremal black brane configurations in string theory and M-theory entail a set of unstable
thermodynamic state-space hypersurfaces. At the zero Hawking temperature, such a limiting
characterization naturally leads to the question of an ensemble of equilibrium microstates
of the extremal black holes and thus the existence of thermodynamic state-space geometry.

Similar explorations exist in determining the role of the thermodynamic fluctuations in finite
parameter Hawking radiating black holes with and without the generalized quantum gravity corrections.
For the Hawking radiating black holes, such an investigation characterizes the intrinsic geometric
description for the quantum statistical physics \cite{ bbt,ZCZ,BonoraCvitan}. Following Ruppenier's
argument, one can take an account of the fact that the zero scalar curvature indicates certain bits
of information on the event horizon fluctuating independently of each other, while the diverging
scalar curvature signals a phase transition indicating highly correlated pixels of the informations.
Fundamentally, Bekenstein \cite{Bekenstein} has introduced an elegant picture for the quantization
of the event horizon area of the black hole, being defined in terms of Planck areas, since a decade.
This led the limiting thermodynamic consideration of finite parameter Hawking radiating configurations
and thus the parametric pair correlations and global statistical correlations. Such an issue intrinsically
serves the motivation for the quantum gravity corrected limiting thermodynamic geometric configurations.

Following Widom's \cite{Widom} initiation of the theory of critical points and positivity of the specific
heat capacities, Refs.\cite{BNTBull,bnt} have interestingly shown that the thermodynamic notions in general
requires the positivity of the principle minors of the determinant of the metric tensor on the state-space
manifold. The global properties of the state-space configurations are revealed from the geometric invariants
on the associated state-space manifolds. From the gravitational aspects of the string theory \cite{0508023,0107119,07073601v2},
one finds that the limiting zero temperature thermodynamic configurations arise from the AdS/ CFT correspondence.
The thermodynamic interpretation of the macroscopic degeneracy may be formally attempted through the partition
function in the grand canonical ensemble involving summation over the chemical potentials. This leads to the
fact that an ensemble of liquid droplets or random shaped fuzzballs pertain a well-defined, non-degenerate,
regular and curved intrinsic thermodynamic surfaces \cite{SAMpaper}.

The origin of the gravitational thermodynamics comes with the
existence of a non-zero thermodynamic curvature, under the coarse
graining mechanism of alike ``quantum information geometry'',
associated with the wave functions of underlying BPS black holes.
Such an intrinsic characterization is highly non-trivial and
interesting in it's own, leading to an exact microscopic
comprehension of 1/2-BPS black holes. Interestingly, the
developments do not stops here, in fact they continue with (i)
rotating spherical horizon black holes \cite{RotBH} in four and
higher spacetime dimensions, (ii) non-spherical horizon topology
black stings and black rings \cite{BSBR,RE} in five spacetime
dimensions, (iii) vacuum fluctuations causing generalized
uncertainty corrections \cite{bntgup}, (iv) plasma-balls in large
N gauge theories \cite{shiraz1,shiraz2}, (v) distribution
functions \cite{CKR2}, associated equations of state of the high
temperature quarks and gluons viscosity \cite{CKR1}, (vi)
thermodynamic properties of QGP in relativistic heavy ion
collisions \cite{CKR3} and (vii) thermodynamic geometric aspects
of the quasi-particle Hot QCDs \cite{bntsbvc, bntsbvc1, bntsbvc2}.

Motivated from such a diverse physical considerations, we herewith intend for the experimental perspectives
of the intrinsic thermodynamic geometry. Along with the above excitements, we explore the modern role
of the thermodynamic geometry to the physical understanding of underlying evolutions, local and global
thermodynamic correlations and possible phase transitions in the due course of the thin film depletion.
To be of interests of the modern experiments, it is worth pointing out that the present designe shares the
viewpoints with an ensemble of finitely many excited non-empty $d$ and $f$-orbitals. Thus, the present paper
examines the intrinsic geometric properties of the thin film depletion layer formation. With the notion
of the adiabatic local evolutions, the random fluctuations in an underlying statistical ensemble offers a
non-linear globally correlated thermodynamic configuration. The evolution parameters, \textit{viz.},
electric charge and depletion mass, describing the fluctuations in the underlying statistical ensemble,
form the coordinate charts on the thermodynamic manifold. The associated scalar curvature determines
the global behavior of the correlation in the system.

In due course of the thin film layer formation, we analyze mathematical nature of the local heat capacities,
global stability and global correlations under the Gaussian fluctuations of the electric charge and mass which
evolve at each state of the respective embeddings. From the definition of the intrinsic Riemannian geometry
and differential topology \cite{Carmo,Bloch}, the present analysis offers an appropriate useful design for coating
a desired thin film material. The quality of the coated product is thus  geometrically optimized with an intrinsically
fine-tuned parametrization of the electric color and mass of the material. To be specific, the present paper
explores the quality of the thin layer depletion, and thus offers an appropriate design for an illusive,
stylish, desired shape, low economic cost, quality-looking products at an affordable price. Form the perspective
of the industrial and daily-life applications, the present exposition anticipates the most prominent gift
of the thermodynamic geometry.

Following this procedure, we consider the statistical theory of the thin film layer formation
with an ensemble of nano-particle depletion. From the mathematical perspective of the intrinsic
thermodynamic geometry, the depleting particles could be positive charges, negative charges,
ions, or a set of other particles, such as electrons, positrons, or any other, if any.
During the thin film depletions, we consider that all the charges are quantized in the units of
electron charge: $\vert e \vert$, and the masses quantized in the units of atomic mass units (AMU): $m$.
Thus, any physical particle carrying an effective electric charge $Q$ and effective mass $M$ can be
described by the two dimensionless parameters, $\{ x, y \}$. For the purpose of the subsequent analysis,
these dimensionless parameters are defined as
\begin{eqnarray} \label{parameters}
x&=& Q/ \vert e \vert \nonumber \\
y&=& M/m
\end{eqnarray}
Notice further that $(\vert e \vert ,m)$ is the pair of experimentally observable elementary electric
charge and elementary AMU, below which present daily-life appliances are of the least importance.
Interestingly, these scaling are the consequences of the Millikan's oil drop experiment and the
Faraday's electrolysis experiment. With this brief physical motivation, we explore the possibility
of the thermodynamic geometry at the present day experiments, in the subsequent sections of the paper.
This would offer the perspective applications of the thermodynamic geometry.

The rest of paper is organized as follow. The section 2 motivates
the study of the small fluctuations, under the depletion layer
formation. Thereby, we set-up our model in the section 3. In
section 4, we offer the specific depletions, and offer the
motivations for the uniform, linear and generically smooth
coatings. In section 5, we describe the possible nature of the
thermodynamic fluctuations over an ensemble of electric charge and
mass, for the case of invertible evolutions. In section 6, we
introduce the notion of intrinsic correlations among the sequence
of charge and mass, and thus an exposition to thermodynamic
geometry, under the Gaussian evolutions. In section 7, we analyze
the stability of the canonical ensemble under the statistical
fluctuations on the thermodynamic surface of the charge and mass.
In section 8, we define the tangent manifold and associated
thermodynamic connection functions. In section 9, we analyze the
global nature of the thermodynamic correlations and possible
phase-transitions. Finally, section 10 contains a set concluding
issues arising from the consideration of the thermodynamic
geometry, offering an outlook for the daily-life experiments and
associated physical implications.
\section{Small Fluctuations under Depletion}
In this section, we consider a nano depletion layer coating of
thickness $l_t$ and coating length $l_L$. Note in the case of the
circular and cylindrical coatings, which are often demanded in the
daily-life applications, $l_t$ would denote the shell of the
thickness $\Delta r:= r_2-r_1 $, where $r_1, r_2$ are the radii of
the inner and outer circles and $l_L$ respectively denotes the
perimeter of the circle or the periphery of the cylinder, as per
the consideration. Thus, these film coatings can be used in
building an illusive low cost perspective designs of the gold,
diamond and platinum and their associated industrial interests.

For a set of chosen materials to be coated over some low coast
material (such as silica), the coating is said to be well-defined
over the local nano-layer formation, if a sequence of the
depleting charge $\{ Q_i \}_{i=0}^N$ and a sequence of the
depleting masses $\{ M_i \}_{i=0}^N$ remain dense over the each
infinitesimal adiabatic evolutions. For the local thermodynamic
correlations, the precise definition of an appropriate density is
described in the section 6.

In the sense of the local function theory, we may express the
mechanism of the thin layer formation as a sequence of the charges
and masses at each stage of the depletion, forming the respective
layers as a set of interval of non-zero widths. At each stage of
an ensemble of nearly equilibrium processes, the charge and mass
$(Q_c, M_c)$ become non-trivially correlated, and thus they can be
depicted as the following expressions
\begin{eqnarray}
Q_c&=& Q(1+ f_Q(x,y)) \nonumber \\
M_c&=& M(1+ f_M(x,y))
\end{eqnarray}
The concerned real embeddings of the charge and mass are defined
as
\begin{eqnarray}
f_Q&:& \mathcal{M}_Q\rightarrow R_Q \nonumber \\
f_M&:& \mathcal{M}_M\rightarrow R_M
\end{eqnarray}
where the domains of $(f_Q,f_M)$ give the allowed values of the
coated $Q$ and $M$, while the ranges of $(f_Q,f_M)$ give the
allowed order of the charge and mass fluctuations over the coating
of the desired nano-layer depletion. Further, the symbols
$\mathcal{M}_Q$ and $\mathcal{M}_M$ denote the respective definite
regions of the coated material with the fixed mass and fixed
electric charge.
\begin{center}\begin{figure}
\hspace*{2.5cm}\vspace*{-2.0cm}
\includegraphics[width=8.0cm,angle=0]{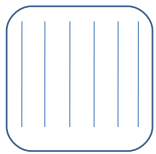}
\caption{A metal depletion as function of the mass $M$ and
electric charge $Q$, describing the depletions of an ensemble of
evolving statistical systems with a constant electric charge
$Q_C$.} \vspace*{-0.2cm}
\end{figure}\end{center}
\begin{center}\begin{figure}
\hspace*{2.5cm}\vspace*{-2.0cm}
\includegraphics[width=8.0cm,angle=0]{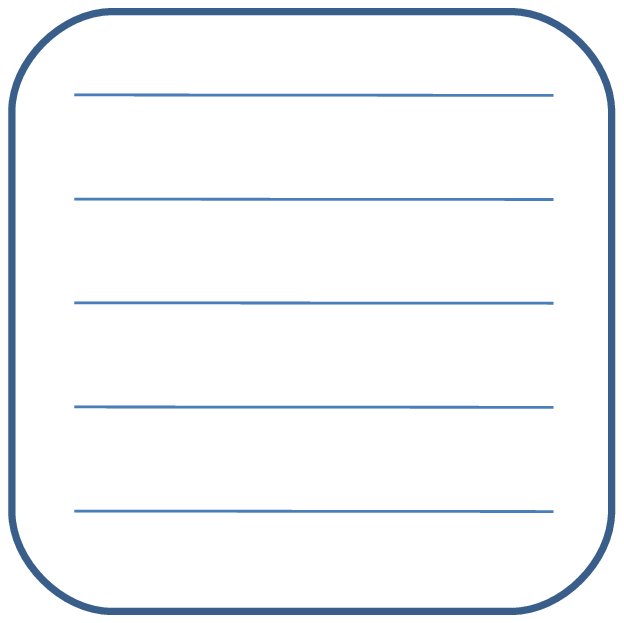}
\caption{A metal depletion as function of the mass $M$ and
electric charge $Q$, describing the depletions of an ensemble of
evolving statistical systems with a constant depleting mass
$M_C$.} \vspace*{-0.2cm}
\end{figure}\end{center}
\begin{center}\begin{figure}
\hspace*{2.5cm}\vspace*{-2.0cm}
\includegraphics[width=8.0cm,angle=0]{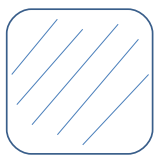}
\caption{A metal depletion as function of the mass $M$ and
electric charge $Q$, describing the depletions of an ensemble of
evolving statistical systems with linearly dependent electric
charge and mass, \textit{i.e.}, $M\sim Q$.} \vspace*{-0.2cm}
\end{figure}\end{center}
\begin{center}\begin{figure}
\hspace*{2.5cm} \vspace*{-2.0cm}
\includegraphics[width=8.0cm,angle=0]{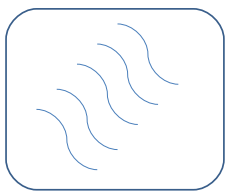}
\caption{General metal depletion as function of the mass $M$ and
electric charge and $Q$, describing the depletions of an ensemble
of evolving statistical systems with an arbitrary spike of the
electric charge and mass: $\{M,Q\}$.} \vspace*{-0.2cm}
\end{figure}\end{center}

The associated experimental characterizations have been respectively shown in the Fig.(1) and Fig.(2).
Herewith, it is an important case of the coating, when the charge and mass are deposited at the same rate.
This configuration has been depicted in the Fig.(3). Finally, the most general cases is the consideration
of an ensemble of depletions, when the electric charge and mass are unrestricted. Such a thin layer
formation is an intrinsically non-trivial configuration. Generically, such an ensemble of diagrams
may be depicted as a set of random spikes. A systematic ensemble could be the case of the Fig.(4).

From the perspective of the equilibrium, meta-equilibrium, quasi-equilibrium and semi-equilibrium
thin layer depletions, the local thermodynamic correlations, including all possible processes of
the interests, may be expressed as the following two composition maps
\begin{eqnarray} \label{jointembedding}
f_{QM}&:=& f_Q \circ f_M: \mathcal{M}_{MQ}\rightarrow R_{QM} \nonumber \\
f_{MQ}&:=& f_M \circ f_Q: \mathcal{M}_{QM}\rightarrow R_{MQ}
\end{eqnarray}
The process of depletion is said to be well-defined and experimentally feasible,
if the composition operation $\circ$ satisfies the following mapping property
\begin{eqnarray}\label{feasible}
f_{QM}=f_{MQ}
\end{eqnarray}
\section{Set-up of the Model}
In the present consideration, the electric charge and mass maps associated with the nano-layer depletions
are physically required to have the following boundary properties
\begin{eqnarray}\label{setup}
\Vert f_Q \Vert &<& L_Q \nonumber \\
\Vert f_M \Vert &<& L_M,
\end{eqnarray}
where the bounds $L_Q$ and $L_M$ give the maximum fluctuations in the electric charge $Q$ and mass $M$.
The composition characterizations of the thin layer depletion are justified with the following
geometric considerations
\begin{eqnarray}
dom(f_{QM})&=& dom(f_{Q}) \cup dom(f_{M}) \nonumber \\
rang(f_{QM})&=& rang(f_{Q}) \cup rang(f_{M})
\end{eqnarray}
In is worth mentioning that the range of such depletions is a set of all possible output values of $f_{QM}$.
This may be defined as the following set of standard embeddings $\{ f_{QM}(x,y): x,y \in M_{QM} \}$.
Thus, the range of $f_{QM}$ could in principle be taken as the same set as the codomain, or a proper subset
of the above standard embeddings. In general, it is designed to be smaller than the codomain, unless the map
$f_{QM}$ is taken to be a surjective coating function.

This outlines the maximum possible domain and the maximum possible range of the electric charge and mass depletions,
when either both of them or one of them fluctuate.
\section{Experimental Set-up}
After illustrating the joint fluctuations as the composition class mappings, we now confront
with the general characterization, which be in particular need not the standard compositions
of the product type. Nonetheless, the above characterization is feasible, if the $f_{QM}$ follows
the above mentioned diagrams, \textit{viz.}, Fig.(1), Fig.(2), Fig.(3) and Fig.(4). Specifically,
such characterizations offer a class of uniform, linear and generic smooth thin film coatings.

As mentioned earlier, we do not only consider the case of the uniform and linear coatings, in specific.
But, we may directly explore the general case of having a smooth class coatings. After the thermal equilibration,
the statistical system reaches the desired thin film equilibrium limit of an interest, \textit{viz.}, Fig.(4).
In this case, it is worth mentioning that both the parameters $\{Q, M\}$ fluctuate independently. Nevertheless,
the linear coating holds locally with the consideration of $\mathcal{M}_{MQ}= \cup_{\alpha}U_{\alpha}$.
Here, $U_{\alpha}$ are an ensemble of open sets on which the depletion of $Q$ and $M$ is desired to be
globally uniform and as smooth as possible. Such a characterization is required in order to have an
illusive high quality product with a relatively low economic input.
\section{Invertible Thermodynamic Evolutions}
We now describe, what could be the possible nature of the thermodynamic fluctuations over the
electric charge and depleting mass $\{Q, M\}$. Considering the present day's requirements,
we may assume physically that the fluctuations under consideration evolve slowly, and in particular
at the infinitesimal scales such as the nano-scale, they take an adiabatic path. Whilst,
there could be finitely many possible global jumps in the system, while the process of the adiabatic depletion
is going-on on the coatant metal frame of the desired shape and size.

Typically, the present day's experiments are interested in the thin film metal depletion of an interval of
$(nm,\ \mu m)$. It is herewith worth mentioning that the possible global jumps are expected to be of an order
of the thickness of the interface between the two phases. The thickness of the phase transition is expected
to remain finite, except at the critical point(s) \cite{RuppPRA44,Arcioni}. The thermodynamic fluctuation theory further
shows that such transitions occur only when the limiting system becomes ill-defined. According to Widom \cite{Widom},
such an instance precisely occurs (i) at the critical points of the system and (ii) along the spinodal curves.

To have a definite invertible movement in the space of $Q$ and $M$, we require that the Jacobian of the
transformation $(x,y) \rightarrow (x^{\prime}, y^{\prime})$ remains non-zero, as the minimal algebraic
polynomial \cite{Yu}. In this concern, the experimentations of the interest must have the following
well-defined movement characterization
\begin{eqnarray}
J(\left (\begin{array}{r}
   (x, y) \\
   (x^{\prime}, y^{\prime}) \\
\end{array} \right))=
\left (\begin{array}{rr}
  \frac{\partial x^{\prime}}{ \partial x} & \frac{\partial x^{\prime}}{ \partial y} \\
     \frac{\partial y^{\prime}}{ \partial x} & \frac{\partial y^{\prime}}{ \partial y} \\
\end{array} \right)\neq 0
\end{eqnarray}
It is worth mentioning that the conditions of having a vanishing Jacobian system,
leads to an irreversible thermodynamic move, and thus it makes the system a non-adiabatic.
Such processes are beyond the scope of present day's daily-life applications.
One may take an account of such movements with a little complication of the non-Markovian moves
\cite{Markov, Gelfand}, requiring an extension of the limit theorems of the standard probability theory.
The present paper do not consider such issues here because they are far from the scope
of the present experiments. Furthermore, these notions on their own need a separate treatment.
Herewith, we shall leave these issues for the future exploration of the present initiation.
\section{Local Thermodynamic Correlations} \label{ltc}
After introducing the electric charge and mass depositing under the thin layer formation, an
appropriate task would now be to introduce the statistical notion arising from the respective
sequence of the electric charge $\{ Q_i \}_{i=0}^N$ and the depleting mass $\{ M_i \}_{i=0}^N$.

Let us consider the most general invertible fluctuations to be
almost everywhere dense over the space of evolution functions,
whose basis vectors are linear combinations of the embeddings
$f_Q$, and $f_M$. Then, the set-up of the present model as defined
in the Eqn.\ref{setup} implies under the aforementioned operation
$\circ$ that the system is at least $L^1$ stable over the
$\mathcal{M}_{QM}$. Nevertheless, this condition is not sufficient
for the stability of the underlying joint ensemble $\{ (Q_i,M_i)
\}_{i=0}^N$, as the physical probability measure. To accomplish
the semi-classical thermodynamic stability of the evolutions, we
require that the adiabatic approximation holds, at least in the
piece-wise evolutions of each local thermodynamic ensemble of
states. Following the standard notion of the quantum physics, we
may thus demand that $\{ (Q_i,M_i) \}_{i=0}^N$ be $L^2$-dense.

From the sense of modern function theory,
the notion of such a density is required because of the volume measure on $\mathcal{M}_{QM}$,
so that one can examine the appropriate class of the probability measures over the distributions
of the electric charge and the mass. To simplify the picture, we wish to work in the quadratic limit,
and henceforth we consider the Gaussian probability measure to be a good approximation. Furthermore,
in order to write the subsequent quantities covariantly, let us introduce the following relabeling
of the dimensionless constants $(x,y)=(x_1,x_2)$. In the present case of the thin film metal depletion
with an ensemble of identical electric charges and depleting masses, one finds that the Gaussian probability
distribution reduces to the following form
\begin{eqnarray}
 P( x^1, x^2)= A\ exp(-\frac{1}{2} g_{ij}\Delta x^i \Delta x^j)
\end{eqnarray}
With respect to an arbitrarily chosen thermodynamic origin $\{x^i_0\} \in \mathcal{M}_{QM}$,
the relative coordinates $\Delta x^i$ are defined as $\Delta x^i:= x^i- x^i_0$.
Taking the standard product measure normalization
\begin{eqnarray}
\int dx^1 dx^2  P( x^1, x^2)= 1,
\end{eqnarray}
we find that the Gaussian probability distribution reduces to
\begin{eqnarray}
 P( x^1, x^2)= \frac{\sqrt{\Vert g \Vert}}{(2 \pi)} exp(-\frac{1}{2} g_{ij}\Delta x^i \Delta x^j),
\end{eqnarray}
where $\Vert g \Vert$ is the determinant of the metric tensor.
In subsequent analysis, we shall set $x^i_0=0$. Thus, we see that the local thermodynamic correlation are achieved,
when the joint probability distribution of an ensemble of the electric charges and the depleting masses $\{ (Q_i,M_i) \}_{i=0}^N$
approaches to an equilibrium thermodynamic configuration.

In the limit, when we take an account of the Gaussian fluctuation as the composite
$f_{QM}$, the local thermodynamic correlations are allowed to go-on over the entire system.
Such objects may thus be defined via the joint embeddings Eqn.\ref{jointembedding},
satisfying the feasibility condition Eqn.\ref{feasible}. In particular, the feasible correlations,
when considered as the Hessian matrix of $f_{QM}$, form local metric structure over the $\mathcal{M}_{QM}$.
From the perspective of the optimal control problem on a manifold of interest \cite{Zelikin},
the corresponding components of the metric tensor reduces as
\begin{eqnarray} \label{metric}
g_{ij}= k H_{ij}(x_1,x_2); \ \vert x_1 \vert, \vert x_2 \vert \in (0,l),
\end{eqnarray}
where $l= max(l_Q, l_M)$ and the chosen sign $k$ is introduced in Eqn(\ref{metric}) in order
to ensure positivity of the metric tensor $g_{ij}$. Physically, it should be noted that the system
should have $\vert l \vert < \{ l_t, l_L \}$. In the above representation of $f_{QM}$, it turns out that
the components of the Hessian matrix as the function of the charge and depleting mass may be expressed as
\begin{equation}
\vert H_{ij} \vert:= \frac{\partial^2 f_{QM}(\vec{x})}{\partial x^i \partial x^j}
\end{equation}
It follows from the above outset how the thermodynamic geometry is employed to describe
the fluctuating blob configurations of the effective electric charge and effective mass.
Under each infinitesimal depletions, the allowed charge-charge self-correlations are
depicted as the set \begin{eqnarray} \{ g_{x_1x_1} \vert \ x_1 \ll Q/e \} \end{eqnarray}
The allowed mass-mass self-correlations are defined as the set
\begin{eqnarray} \{ g_{x_2x_2} \vert \ x_2 \ll M/m \} \end{eqnarray}
Finally, the charge-mass inter-correlations are simply depicted as the set
\begin{eqnarray} \{ g_{x_1x_2} \vert \ x_1 \ll Q/e, \ x_2 \ll M/m \} \end{eqnarray}
Due to a small intersection over the domains of the mass and charge, we observe that
the inter-correlations are expected to be much smaller than the pure charge and
mass self-correlations. It follows further from the standard fact that the product
of the variables $x_1$ and $x_2$ defines the allowed area of the interest on $\mathcal{M}_{QM}$.

\section{Ensemble Stability Condition}

The stability of the statistical fluctuations over $\mathcal{M}_{QM}$ can be
determined with respect to the local fluctuations in $Q$ and $M$. Such a condition
is ensured, whenever $g_{x_1x_1}$ and $g_{x_2x_2}$ as the charge-charge and mass-mass
heat capacities remain positive on $\mathcal{M}_{QM}$. If one of the variable fluctuates
much larger than the other, then the larger fluctuations should be positive in order
to have a locally stable thermodynamic configuration. It is worth mentioning that
the charge-charge and mass-mass self correlations are known as the positive heat capacities.
The stability of the statistical system holds along a chosen direction, if the other
variable remain intact under the thermodynamic fluctuations.

Whenever there exists a non-zero finite inter-correlation involving both of the
directions on $\mathcal{M}_{QM}$, then the thermodynamic fluctuations is said to be
system stable, if the determinant of the metric tensor
\begin{eqnarray}
det (g_{ij}):= \Vert g \Vert= g_{x_1x_1} g_{x_2x_2}-  g_{x_1x_2}^2 >0
\end{eqnarray}
The vanishing of $\Vert g \Vert$ leads to the unstable large
thermodynamic fluctuations. In such cases the global configuration
has an ill-defined surface form, and thus the possibility of a
leading non-orientable $\mathcal{M}_{QM}$.It is worth mentioning
that these issues are certainly in their own, but at this moment
they are least interesting from the prospectiveness of
experimental affairs of the intrinsic thermodynamic geometry.

\section{Thermodynamic Connection Functions}
At this stage, we wish to consider those experimental observations
which are of global nature and which could be arising from
topological considerations of the intrinsic thermodynamic
geometry. The topological defects \cite{Nakahara} of the present
interest, are a class of stable objects against small
perturbations, which do not decay or become undone or de-tangled,
because there exists no continuous transformation that can
homotopically map them to a uniform solution. Thus, to compute
such a globally invariant quantity, we need to define the
Christoffel connections on $\mathcal{M}_{QM}$. The Christoffel
symbols \cite{waldGR} are most typically defined in a coordinate
basis, which is the convention to be followed here. It follows
further, from the definition of the dimensionless quantities
$\{x_1, x_2 \}$, that they form a local coordinate system on the
$\mathcal{M}_{QM}$. The definition of the directional derivative
along $x_i$ gives a pair of tangent vectors
\begin{eqnarray}
e_i= \frac{\partial}{ \partial x^i},\ i= 1,2
\end{eqnarray}
Locally, this defines a complete set of basis vectors on the
tangent space $T \mathcal{M}_{QM}$, at each point $ p \in
\mathcal{M}_{QM}$. Given the composite map $f_{QM} \in
\mathcal{M}_{QM}$, the Christoffel symbols $\Gamma^k_{ij}$ can be
defined as the unique coefficients such that the following
transformations
\begin{eqnarray}
\nabla_i e_j= \Gamma^k_{ij}e_k
\end{eqnarray}
hold, where $\nabla_i$ is understood as the Levi-Civita connection
on the charge-mass manifold $\mathcal{M}_{QM}$, which is taken in
the coordinate direction $e_i$. The Christoffel symbols can be
further derived from the vanishing condition of the Hessian matrix
of the composite map $f_{QM}$. This follows from the fact that
$H_{ik}(x_1,x_2)$ defines the notion of the covariant derivative
with respect to the metric tensor $g_{ik}$. In this definition, we
consider that $g_{ij}$ has the standard meaning, viz., it can be
defined as an inner product $g(\frac{\partial}{\partial x^i},
\frac{\partial}{\partial x^j})$ on the tangent space
$T(\mathcal{M}_{QM}) \times T(\mathcal{M}_{QM})$ with the
following determinant of the metric tensor
\begin{eqnarray}
g(x):= \Vert g_{ij} \Vert
\end{eqnarray}
As mentioned in the foregoing section, the determinant of the
metric tensor $g(x)$ is regarded as the determinant of the
corresponding matrix $[g_{ij}]_{2 \times 2}$. Thus, for a given
charge-mass manifold $\mathcal{M}_{QM}$, we shall think that the
Christoffel symbols can be expressed as a function of the metric
tensor. Explicitly, such a consideration leads to the following
formula
\begin{eqnarray}
\Gamma^i_{jk} = {1 \over 2} g^{im} (g_{mj,k} + g_{mk,j} -
g_{jk,m}),
\end{eqnarray}
where $[g^{jk}]$ is the inverse of the matrix $[g_{jk}]$,
satisfying the identity $g^{j i} g_{i k}= \delta^j_k\ $.

Interestingly, the Christoffel symbols are written with the tensor
indices, however, it is not difficult to show, from the
perspective of coordinate transformations, that they do not belong
to the tensor family. Although, the Christoffel symbols are useful
in defining tensors, but they are themselves examples of the
non-tensors. An immediate example of such a construction is the
matter of the next section.

\section{Global Thermodynamic Correlations}

As mentioned in the foregoing section, we shall setup the notion
of the global correlation, about an equilibrium. In the next
section, we shall offer a numerical proposal for the depletion of
the electric charge and mass. For given charge and mass maps, this
proposal turns out to be minimal, viz., the entire configuration
can locally be considered as a well-defined and non-interacting
statistical system. Before doing so, let us consider a vector in
$\mathcal{M}_{QM}$, then we find, when it is parallel transported
around an arbitrary loop in $\mathcal{M}_{QM}$, that it does not
return to its original position. This could simply be taken into
account by the holonomy \cite{waldGR} of charge-mass manifold
$\mathcal{M}_{QM}$. Specifically, the Riemann-Christoffel
curvature tensor measures the holonomy failure on
$\mathcal{M}_{QM}$. Such a consideration defines the possibility
of a non-trivial geometric depletion of the thin film nano-slab of
experimental interest.

To see the deviation, let $x_t$ be a curve in  $\mathcal{M}_{QM}$.
Denoting $ \beta_t: T_{x_0}\mathcal{M}_{QM} \rightarrow T_{x_t}
\mathcal{M}_{QM}$ as the parallel transport map along $x_t$, then
the covariant derivative takes the following form
\begin{eqnarray}
\nabla_{\dot{x}_0} X_2 = \lim_{h\to 0}
\frac{1}{h}\left(X_{2_{x_0}}-\beta^{-1}_h(X_{2_{x_h}})\right) =
\left.\frac{d}{dt}(\beta_{x_t}X_2)\right|_{t=0}
\end{eqnarray}
for each vector field $X_2$ defined along the curve $x_t$. In
order to explicitly compute the deviation, let $(X_1,X_2)$ be a
pair of commuting vector fields, then each of these fields
generate a pair of one-parameter groups of diffeomorphisms in a
neighborhood of $x_0 \in \mathcal{M}_{QM}$. Denoting
$\beta_{t_{X_1}}$ and $\beta_{t_{X_2}}$ respectively the parallel
transports along the flows of $X_1$ and $X_2$ for a finite time $t
\in (0, \infty)$, then the parallel transport of a vector $X_3 \in
T_{x_0} \mathcal{M}_{QM}$ around the quadrilateral of the sides
$\{tX_2$, $sX_1$, $-tX_2$, $-sX_1\}$ is given by the following
composition
\begin{eqnarray}
\beta_{sX_1}^{-1}\beta_{tX_2}^{-1}\beta_{sX_1}\beta_{tX_2}X_3
\end{eqnarray}
This measures the holonomy failure of the vector field $X_3$,
under its parallel transport to the original position on $T_{x_0}
\mathcal{M}_{QM}$. Now, if we shrink the loops to a point, viz.,
taking the limit $s, t \rightarrow 0$, then the infinitesimal
description of the above deviation is given by
\begin{eqnarray}
\left.\frac{d}{ds}\frac{d}{dt}\beta_{sX_1}^{-1}\beta_{tX_2}^{-1}
\beta_{sX_1}\beta_{tX_2} X_3 \right|_{s=t=0} = (\nabla_{X_1}
\nabla_{X_2} - \nabla_{X_2} \nabla_{X_1}) X_3 = R(X_1,X_2) X_3
\end{eqnarray}
where $R(X_1,X_2)$ is the Riemann curvature tensor on
$\mathcal{M}_{QM}$. Notice that the dimension of
$\mathcal{M}_{QM}$ is a two dimensional manifold, and thus
$R(X_1,X_2)\equiv R_{ijkl}$ has only one non-trivial component.
This component of the Riemann curvature tensor takes the following
form
\begin{eqnarray}
R_{1212}= \frac{N}{D},
\end{eqnarray}
where
\begin{eqnarray}
N&:=& S_{22}S_{111}S_{122} + S_{12}S_{112}S_{122} \nonumber \\ &&+
S_{11}S_{112}S_{222} -S_{12}S_{111}S_{222} \nonumber \\ &&-
S_{11}S_{122}^2- S_{22}S_{112}^2
\end{eqnarray}
and
\begin{eqnarray}
D:=  (S_{11}S_{22}- S_{12}^2)^{2}
\end{eqnarray}
Here, the subscripts on $S(X_1, X_2)$ denote the corresponding
partial derivatives pertaining the Hessian matrix $k
H_{ij}(x_1,x_2)$. For a smooth $f_{QM} \in \mathcal{M}_{QM}$, it
turns out that the dual variables $X_i$ are defined via the
Legendre transformation, viz., $X^i:= \frac{\partial
f_{QM}(x)}{\partial x_i}$. For any two-dimensional
$\mathcal{M}_{QM}$, the Bianchi identities imply that the Riemann
tensor can be expressed as a function of the coordinates and the
metric tensor
\begin{eqnarray} \label{Riemann}
 R_{abcd}^{}=K(g_{ac}g_{db}- g_{ad}g_{cb}) \,
\end{eqnarray}
where $g_{ab}$ is the metric tensor and $K(x_1,x_2)$ is a function
called the Gaussian curvature of $\mathcal{M}_{QM}$. In the
Eqn.(\ref{Riemann}), the indices $a$, $b$, $c$ and $d$ take the
values either 1 or 2. It is well-known \cite{Carmo,waldGR} that
the Gaussian curvature coincides with the sectional curvature of
the charge-mass surface, and it is exactly the half of the scalar
curvature of $\mathcal{M}_{QM}$. Consequently, the Ricci curvature
tensor of the charge-mass surface takes the following form
\begin{eqnarray}
 R_{ij} = {R^k}_{ikj} = Kg_{ab}. \,
\end{eqnarray}
Given the determinant of the metric tensor and Riemann-Christoffel
curvature tensor $R_{1212}$, the Ricci scalar curvature of the
corresponding two dimensional thermodynamic intrinsic manifold
$(\mathcal{M}_{QM}(R),g)$ can be expressed by the following
formula
\begin{eqnarray}
R(x_1,x_2)=\frac{2}{\Vert g \Vert}R_{1212}(x_1,x_2)
\end{eqnarray}
It is worth mentioning further that $\mathcal{M}_{QM}$ is a space
form, if its sectional curvature coincides with the constant $K$,
and then the Riemann tensor is of the form of the
Eqn.(\ref{Riemann}). Thus, it is straightforward to analyze the
nature of the thin film layer formation, desired global coatings,
canonical correlations, and possible phase transitions, as the
thin film depletion involve only finitely many codings of the
electric charge and mass.
\section{Experimental Verification}
In the present section, we shall offer a proposal for the
experimental test of the local and global thermodynamic
correlations. For a given equilibrium value of the electric charge
and mass, this section provides a numerical code such that the
desired depletion of the material remains a homogeneous phase.
Locally, this requires a fixation of the concerned scalings of the
charge and mass.

For $x \in \mathcal{M}_{QM}$, let $L:= \{x_i\}_{i=0}^n$ be a
finite coding sequence, such that the finite difference
$x_i-x_{i-1}=h$ defines an interval on $L$. Since, the step size
$h$ remains the same at all evolutions, thus $L$ is an evenly
spaced lattice. In order to illustrate the model, let us first
consider the lattice $L$ to be one dimensional. Let $f(x_i)=f_i$
be a numerical sequence corresponding to the respective values of
$x_i \in L$. Then, the first derivative of $f$ is given by
\begin{eqnarray} \label{numerical}
f_i^{\prime}=\frac{f_i-f_{i-1}}{h}
\end{eqnarray}
In order to examine the numerical nature of the local and global
thermodynamic correlations, we need the first few derivatives of
the composite embedding of the charge and mass maps. Herewith, we
find that the higher derivatives $f_i^{\prime \prime},
f_i^{\prime\prime\prime}$ take the following forms
\begin{eqnarray}
f_i^{\prime \prime}=\frac{f_i- 2f_{i-1}+ f_{i-2}}{h^2}
\end{eqnarray}
\begin{eqnarray}
f_i^{\prime \prime \prime}=\frac{f_i- 3f_{i-1}+ 3f_{i-2}-
f_{i-3}}{h^3}
\end{eqnarray}
Observing that $f^{n}=0$, if $f_i=f \forall i=0,1,2,...,n $, then
the choice of the minimally coupled depletion of the charge and
mass can be offered by the following proposition.

\subsection*{Proposition} Let $\{i\}$ be a collection of points on
$L$, and let $n$ denote the order of the step of the corresponding
depletion. Then, the replacement $f^{(n)}_i:=i^n$ offers the code
for the thermodynamic couplings on the mass-charge surface $
\mathcal{M}_{QM}$.

This code possesses all practical information of the depletion of
the charge and mass. Subsequently, we show that the proposal
becomes minimally coupled, in the limit when the chosen
equilibrium $(x_{10}, x_{20})$ is far separated from the others.
Physically, this means that the local equilibrium $(x_{10},
x_{20})$ is such that the concerned mixed partial derivatives are
evaluated in the limit of their product, e.g. when, as shown
below, the effective scaling be defined as the product of the
scalings along the dimensions $x_1$ and $x_2$.

A physical proof of the proposal can be offered as follows.
Following the definition of the numerical differentiation
Eqn.(\ref{numerical}), we see that the above proposal leads to the
following expressions
\begin{eqnarray}
f_i^{\prime}=\frac{i-(i-1)}{h}=\frac{1}{h}
\end{eqnarray}
\begin{eqnarray}
f_i^{\prime \prime}=\frac{i^2- 2(i-1)^2+
(i-2)^2}{h^2}=\frac{2}{h^2}
\end{eqnarray}
\begin{eqnarray}
f_i^{\prime \prime \prime}=\frac{i^3- 3(i-1)^3+ 3(i-2)^3-
(i-3)^3}{h^3}=\frac{6}{h^3}
\end{eqnarray}
As per the above characterization, we herewith examine the nature
of the stability and non-Euclidian behavior of the thermodynamic
interactions. For such a demonstration, we have already introduced
the notion of the local and ensemble stabilities, the Christoffel
symbol and the Riemann-Christoffel curvature tensor on the surface
$\mathcal{M}_{QM}$. For a given local basis $(x_1,x_2)\in
\mathcal{M}_{QM}$, let us compute the determinant of the metric
tensor and the component of the Riemann-Christoffel curvature
tensor.

Let the evolution of the charge and mass be locally characterized
by the pair $(h_1,h_2)$. Then, in the limit of an ensemble of far
separated equilibria, the determinant of the metric tensor reduces
to the following expression
\begin{eqnarray}
\Vert g \Vert= S_{11} S_{22} -(S_{12})^2= \frac{3}{h_1^2 h_2^2}
\end{eqnarray}
The qualitative behavior of the limiting ensemble stability is
depicted in the Fig.(5). This shows that viability of the code,
with respect to the chosen scales $(h_1,h_2)$.

As mentioned in the previous section, an evaluation of the
(scalar) curvature requires the computation of $N$ and $D$. In
fact, it follows that $D$ can be determined from the $\Vert g
\Vert$. Subsequently, our proposal is proved by showing that the
factor $N$ vanishes identically, in the above mentioned scaling
limit. Explicitly, the vanishing of $N$ can be verified by
considering the previously mentioned factorizations, e.g.,
$S_{112}= (2/h_1^2)(1/h_2)$. This completes the proof of the
proposal.

Herewith, we find that the above proposal is well-defined with
positive heat capacities, viz., $f_i^{\prime \prime}>0$ and a
positive determinant of the metric tensor. Thus, with the present
proposal, the entire evolving configuration of the mass and charge
can be locally considered as a well-defined and non-interacting
statistical system.
\begin{center}\begin{figure}
\includegraphics[width=8.0cm,angle=0]{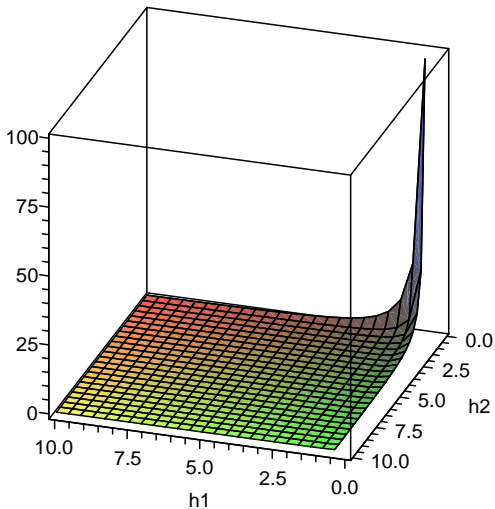}
\caption{Determinant of the metric tensor as a function of the
mass and electric charge scalings $h_1$ and $h_1$, describing the
depletions of an ensemble of evolving statistical systems.} 
\end{figure}\end{center}
\section{Conclusion and Outlook}
In this paper, we have explored experimental perspective of the
thermodynamic geometry. Our analysis is particularly suitable for
the thin film depletion layer formation at a small scale structure
formation under the fluctuations of the deposited electric charge
and the atomic mass. Keeping in mind the nature of finitely many
excited, non-empty $d$/ $f$-orbital heavy materials, we have
investigated the detailed mathematical picture of the intrinsic
thermodynamic geometric and topological characterizations of the
small fluctuation over a depletion layer formation.

Under each adiabatic evolution of the local thermodynamic
macro-states, we have examines the nature of (i) electric charge,
and mass fluctuations under the respective embeddings, (ii)
positivity of the local heat capacities, (iii) global
thermodynamic stability of the canonical ensemble under the
Gaussian fluctuations of a sequence of the electric charges and
depleting masses, and (iv) global charge-mass correlations. Thus,
we have offered a detailed experimental perspective of the
intrinsic thermodynamic geometry.

For any non-degenerate thermodynamic metric tensor and a regular
Gaussian curvature $K(x_1,x_2)$, we have generically shown that
there are no global phase transitions on $\mathcal{M}_{QM}$. On
the other hand, there may exist finitely many critical points,
which are predicted to occur at the roots of the determinant of
the metric tensor. Such a finite critical set may be given as
\begin{eqnarray}
\mathcal{C}:= \{ c_1, c_2, \ldots, c_n \}
\end{eqnarray}
In general dimensions, there may be diverse critical properties of
the $\mathcal{M}_{QM}$. For regular $N(x_1,x_2)$, we notice in the
two dimensions that the global phase transitions only occur
precisely over the set $\mathcal{C}$. This classification thus
effectively confirms notion of Widom’s spoidal curves
\cite{Widom} and the associated global critical phenomena, which
are prone to occur under the metal depletion and it's thin layer
formation.

It is worth mentioning that the order of $R(x_1,x_2)$ determines
the order of the phase transition in the system. Thus, the general
consideration of $\mathcal{M}_{QM}$ makes the function
$N(x_1,x_2)$ more involved. Specifically, when $N(x_1,x_2)$ is a
singular function on a patch of the $\mathcal{M}_{QM}$, then the
global phase transitions can occur, even if the metric tensor
$g_{ij}$ is non-degenerate. In this case, the only requirement to
exist a global phase transition is that both the electric charge
and the mass fluctuations should be finite and non-zero under a
layer formation. From the perspective of numerical analysis, we
have proposed a well-defined numerical code of the lattice
evolutions. The proposed code corresponds to a non-interacting
local statistical system, whenever the chosen equilibrium remains
far separated from the underlying ensemble of the equilibria.

To determine the above notions experimentally or within the scope
of intrinsic geometric model, we need to specify the sample, which
we have described in this paper as the possible local set-up of
our model. For an exact surface modeling, one may choose a fined
shape of certain slab, such as square/ rectangle, circle, ellipse.
It is further possible to chose some quadrilaterals, like
cross-quadrilateral, butterfly quadrilateral, bow-tie
quadrilateral and other skew quadrilaterals. For a near-surface
daily-life modeling, we may add an extra tiny third direction
(having a size of few nano-meter to few micro-meter) to the
present intrinsic geometric surface modeling. In such cases, one
may thus consider a near-surface modeling for the this layer
depletion over a fixed desired image, such as table, thin
cylinder, thin prism, thin pyramid, thin fridge, thin sheeted
stairs, small regulated cone without the tip, thin ellipsoid, and
finally any thin shell of radii $r_1$ and $r_2$ with $r_1 \sim
r_2$. It is needless to mention that the present characterization
holds for any such possible similar pattern formation. In the
sense of the statistical physics \cite{Huang,Landau} and modern
aspects of the functional analysis \cite{Rudin}, the present
exposition offers a microscopic understanding of the thin layer
depletion and pattern formation.

The above classes of the shapes are useful in daily-life
appliances for the composition of an illusive, high-cost, precious
looking objects and the associated materials possibly useful in
perspective decorations. Our method is very desirable in
determining the quality of thin film coating, and thus in
determining the local and the global nature of the coated layer on
a low cost material, \textit{viz.} silica. Following such a
characterizing procedure, one may control the quality of thin
layer depletion. This can be further useful in producing a durable
illusive, stylish, low economic factor, quality-looking products
at a desired economic value. Thus, such an investigation leads to
several industrial importance, offering a class of possible
daily-life appliances, from the application of the intrinsic
thermodynamic geometry.

Apart from the above mentioned considerations, the designed method
is further applicable to arbitrary shape coating's on definite
low-economic-factor frame, \textit{viz.} $\mathcal{M}_{QM}$. At
the desired scale, such a surface of coatants is a randomly
fluctuating surface, which after the equilibration leads to a
desired quality quoted shape, at a large scale. Looking after the
present experimental set-up's and associated daily-life demands,
we may set-up the scale of the experimental thermodynamic geometry
in the order of a few nano-meter to a few micro-meter. Although,
the large scale coating are well-explained from the very out-set
of the present model, however their present daily-life
significance involved are of minor importance, which from the
demand based economic perspective, so it might be herewith
practically of minor noteworthy to emphasize their details.

To achieve the global quality shape thin layer depletion, one only
needs to compute the Gaussian curvature $K(x_1,x_2)$. Thereby, one
may deduce the topological nature of the stability of underlying
structure formation, and global phase-space correlations on
$\mathcal{M}_{QM}$. In summary, the thin layer characterizations
of desired formation can easily be acquired by studying the local
and global properties of fluctuating surfaces, arising under the
depletion of the effective electric charge and depletion mass.

\section*{Acknowledgement}
BNT would like to thank Prof. V. Ravishankar for his support and
encouragement towards the research and higher education. BNT
acknowledges the postdoctoral research fellowship of the
\textit{``INFN, Italy''}.

\end{document}